\documentclass[aps,prl,twocolumn,showpacs,superscriptaddress,groupedaddress,amsmath,nofootinbib]{revtex4-1}  
\usepackage{graphicx}  
\usepackage{dcolumn}   
\usepackage{bm}        
\usepackage{amssymb}
\usepackage{url}

\usepackage{aas_macros}

\begin{document}

\title{Interpretation of the excess of antiparticles within a modified paradigm of galactic cosmic rays}
\author{Ruizhi Yang}\email{ryang@mpi-hd.mpg.de}
\affiliation{%
\ Department of Astronomy, School of Physical Sciences, University of Science and Technology of China, Hefei, Anhui 230026, China\\
\ CAS Key Labrotory for Research in Galaxies and Cosmology,  University of Science and Technology of China, Hefei, Anhui 230026, China\\
\ School of Astronomy and Space Science, University of Science and Technology of China,Hefei, Anhui 230026, China\\
\ Max-Planck-Institut f{\"u}r Kernphysik, P.O. Box 103980, 69029 Heidelberg, Germany\\
\ 
}
\author{Felix Aharonian}\email{felix.aharonian@mpi-hd.mpg.de}
\affiliation{%
\ Dublin Institute for Advanced Studies, 10 Burlington Road, Dublin 2, Ireland \\
\ Max-Planck-Institut f{\"u}r Kernphysik, P.O. Box 103980, 69029 Heidelberg, Germany\\
\ Dublin Institute for Advanced Studies, 31 Fitzwilliam Place, Dublin 2, Ireland\\
MEPHI, Kashirskoe shosse 31, 115409 Moscow, Russia\\
}

\date{Received:  / Accepted: } 

\begin{abstract}
We argue that the anomalously high fluxes of positrons and antiprotons found in cosmic rays (CR) can be satisfactorily explained by introducing two additional elements to the current "standard" paradigm of Galactic CRs.  First, we propose that the antiparticles are effectively produced in interactions of primary CRs with the surrounding gas not only in the interstellar medium (ISM)  but also inside the accelerators.  Secondly, we postulate the existence of two source populations injecting CRs into the ISM  with different, (1) soft (close to $FI \propto E^{-2.3}$) and (2) hard ($FII \propto E^{-1.8}$ or harder),    energy distributions.  Assuming that CRs in  the 2nd population of accelerators accumulate "grammage" of the order of $1 \ \rm g/cm^2$   before their leakage into ISM, we can explain the energy distributions and absolute fluxes of both  positrons and antiprotons, as well as the fluxes of secondary nuclei of the (Li,Be,B) group. The superposition of contributions of two source populations also explains the reported hardening of the spectra of CR protons and nuclei above 200 GV. The 2nd source population accelerating CRs with a rate at the level below 10 percent of the power of the 1st source population, can be responsible for the highest energy protons and nuclei of Galactic CRs up to the ``knee" around  $10^{15} \ \rm eV$.
\end{abstract}
\maketitle

\section{Introduction}
Cosmic Rays (CRs)  consist of primary and secondary components, the latter being the result of interactions of the primary (directly accelerated)  protons and nuclei with the surrounding gas.  According to the current concept of the origin of Galactic CRs, these interactions take place predominantly in the interstellar medium  (ISM). This simple assumption based on the measured content of the secondary light nuclei in  CRs allows several important conclusions regarding the energy-dependent propagation of CRs in the interstellar magnetic fields.  Although in this scenario the positrons and antiprotons appear as unavoidable counterparts of  the secondary nuclei, the measurements by the Pamela satellite \citep{pamelap} and the Alpha Magnetic Spectrometer  
(AMS-02) \cite{ams02posi19} revealed unexpectedly hard energy spectra and significantly enhanced fluxes of positrons and antiprotons.  The reports on the ``excess''  of antiparticles have been received and interpreted by many researchers as a ``smoking gun" of  Dark Matter (see, e.g., \citep{ams02posi19}).    
Such a strong claim in the context of one of the most fundamental objectives of the modern physics and astrophysics requires a careful judgment through the "Occam's razor" principle, i.e., exploration of other, more conventional interpretations of the antiparticle  ``excess''   in CRs \citep{Profumo}.  
While for positrons a possible solution could be the direct contribution from specific "primary"  electron-positron sources (accelerators),  in particular by pulsars,  the interpretation of the reported excess of antiprotons is more problematic. 
Except for the Dark Matter channel, the only realistic possibility is
the antiproton production through interactions of directly accelerated  CRs with the ambient gas. The observed almost energy-independent $e^+/\bar{p}$ ratio (see, e.g.  ref.\cite{Lipari}) seems to be an appealing indication for the intrinsic link between the positrons and antiprotons through the CR related channels.  In other words, the detected flux of antiprotons predicts production of positrons at the level which is close to the reported positron flux and thus leaving not much room for contributions of other sources to the observed positron flux. In particular, the reported antiproton flux significantly limits the fraction of CR positrons  contributed by pulsars and pulsar wind nebulae.   

In summary, the new results on CR positrons and antiprotons highlight the interconnection of the positron "excess"  to the collisions of primary protons and nuclei with the surrounding matter. In this work, we demonstrate that the content of positrons and antiprotons in CRs can be explained within a  modified, two CR source-population paradigm of  Galactic CRs as secondaries produced at interactions of primary CRs with the ambient gas both in the ISM  and inside the CR accelerators. Remarkably, the two source-population concept gives a natural explanation also to the recently discovered result -  the spectral hardening detected in the energy distributions of primary CR protons and nuclei at energies of several hundred GV. 

\section{Production of CR secondaries}

Generally, the spectra of high-energy positrons and antiprotons produced in nuclear reactions mimic the spectrum of parent protons and nuclei.  Later, because of the energy-dependent diffusion, the spectra of positrons and antiprotons become substantially steeper compared to the primary (acceleration) spectrum of  CRs.  Thus,  one can immediately conclude that the positrons and antiprotons detected with hard spectra have not been produced in the ISM.  In this regard, the models proposing production of secondary positrons and antiprotons in a nearby supernova remnant (SNR)  \cite{kachelriess15b,kohri15} seem more attractive.

The option of a single local supernova remnant as a source of  CR positrons and antiprotons has robust limitations; it cannot be located much further than 100~pc and cannot be much older than $10^4$ yr.  Otherwise, the density of CRs would be reduced to a negligible level. The very idea of a nearby local  SNR as the main contributor to the secondary positrons has been proposed while ago  \citep{AAV1}  in the same context of the interpretation of the first (tentative) claim of the anomalously high content of CR positrons  \citep{Chicago}.  
The realization of this model implies effective interaction of directly accelerated protons and nuclei with the ambient gas inside the remnant \citep{AAV1}.  CRs can accumulate significant "grammage" also in the vicinity of SNRs due to the slow propagation caused by self-generation of plasma waves by escaping particles \cite{Malkov13,nava16,dangelo18}. 
A key requirement of this and, in fact, of any other model that assumes production of positrons and antiprotons inside or in the proximity of CR accelerators,  is the hard spectrum of antiparticles injected into the ISM. Therefore, the initial (acceleration) spectrum should also be hard, close to  $E^{-2}$ or harder.   In the ISM, the energy-dependent propagation modifies (makes steeper) the  CR spectrum. Even though, the latter remains harder than $E^{-2.5}$ given that the diffusion coefficient   $D(E) \propto E^\delta$ with $\delta \approx 0.5$  as it follows from the secondary CR data. On the other hand, the measured spectrum of CR protons at low energies is very steep, 
${\rm dN}/{\rm d}E \propto E^{-2.85}$ \cite{ams02proton}.  

Combined with the measured secondary-to-primary ratio of CRs \cite{ams02bc}, one may conclude that the initial spectrum of protons injected into the ISM at energies less than 100 GeV, should be close to $E^{-2.35}$. This implies that, in addition to the {\it first} source population, one should invoke an additional   CR population characterized by significantly harder energy spectrum. The latter demands significantly less injection power than the first (steep) component but, because of the harder spectrum, might dominate at higher energies. Below we show that this assumption can naturally explain the reported hardening of the spectrum of primary CR protons and nuclei above 200~GV.  With an additional assumption of a non-negligible ($\approx 1 \ \rm g/cm^2$) "grammage" accumulated inside the 2nd population sources,  we can explain the fluxes of the secondary products of CRs - positrons, antiprotons, as well as light nuclei of the (Li, Be, B) group.

\section{Results}
 
Under the assumption that all secondaries are produced by primary CRs injected with a power-law spectrum $Q(E)$,  the steady-state distribution of particles in the ISM is   $N(E) \propto Q(E)\tau(E)$. Here $\tau(E)$ is the confinement time of CRs in the Galaxy; it decreases with energy as $\tau(E) \propto  E^{-\delta}$.  Thus, for a power-law injection spectrum, $Q(E) \propto E^{-\gamma_{}}$,  we have $N(E) \propto E^{-(\gamma_{}+\delta)}$.  By ignoring the slight dependence of the inelastic cross-sections on energy above 10~GeV, we assume, as a zeroth order approximation,  that the production spectra of secondaries mimic the steady-state spectrum of primaries in the ISM. Due to the energy-dependent propagation, which obviously should be described by the same diffusion coefficient as  for  primaries, the steady-state spectrum of  secondaries becomes  $N_{\rm sec}(E) \propto Q(E) \tau^{2}(E) \propto  E^{-(\gamma_{}+2\delta)}$. Thus, the  secondary-to-primary ratio monotonically decreases with energy,  $R(E)=N_{\rm sec}/N \propto E^{-\delta}$. This is true for all secondary 
particles except for the very high energy electrons and positrons. Because of the radiative (synchrotron and inverse Compton) cooling, the spectra of secondary electrons and positrons at very high energies become steeper. 

The  secondary-to-primary ratio can be expressed as  (see e.g. ref.\cite{katz10}), 
\begin{equation}
R(E)=\frac{\frac{X_{\rm ISM}(E)}{m_p}S(E)/N_p(E)}{1+\sigma_{\rm t}\frac{X_{\rm ISM}(E)}{m_p}}, 
\end{equation}
where
\begin{equation}
S(E)=\int_E\frac{d\sigma_{\rm p \to s}(E,E')}{dE'}N_p(E')dE'.
\end{equation}
Here E is the particle energy per nucleon, $N_p(E)$ is the primary CR spectrum formed in the ISM, $X_{\rm ISM }$ is the propagation length in $\rm g/cm^2$ ("grammage"), $\sigma_{\rm p \to s}$ is the  differential cross section  of production of the given secondary particle,  $\sigma_{\rm t}$ is the total destruction cross-section of secondaries.  For the spallation reactions,  the energy per nucleon before and after the reaction remains the same (see, e.g., ref.\cite{webber03}). Thus Eq.(1) can be reduced to
\begin{equation}
R(E)=\frac{\frac{X_{\rm ISM}(E)}{m_p}\sigma_{p \to s}}{1+\sigma_{\rm t}\frac{X_{\rm ISM}(E)}{m_p}} . 
\end{equation}
 For the antiproton production cross-sections,  we use the compilations from ref.\cite{winkler18}. The cross-sections of nuclear reaction are taken from ref.\cite{webber03}. For the production of the secondary electrons and positrons,  we use the parametrizations from ref.\cite{yang18}.
For the nuclei spallation reactions, we use the parametrizations of cross-sections from ref.\cite{letaw83}. 

At high energies, the radiative cooling of electrons and positrons becomes an essential factor.  For calculations of fluxes of electrons and positrons, we use the semi-analytical approach developed in ref.\cite{AAV2}.  
As a first-order approximation, we assume a homogeneous distribution of secondary positron and electron sources inside the Galactic plane and a continuous injection and perform integration over the  Galactic plane with a radius of $\sim$ 25~kpc.


The "grammage" accumulated in the ISM is proportional to the confinement time,  
\begin{equation}
 X_{\rm ISM}(E) = c~n_{ism}~m_p~\tau_{}(E), 
\end{equation}
where  $c$ is the speed of light,  and $m_p$ is the  proton mass.
We present  $X_{\rm ISM}(E)$ in the form 
\begin{equation}
X_{\rm ISM}(E)=X_0 (\frac{E}{10~\rm GeV})^{-\delta} \ .
\end{equation}
Note that $X_{\rm ISM}$ depends on rigidity, $R=\frac{AE}{Ze}$, where $A$ and $Z$ are  the atomic and mass number, respectively. For  
protons and antiprotons,  R=E, while for nuclei it differs by a factor of $\approx 2$.  
 Below we limit our consideration  by energies exceeding $30$ GeV  to neglect the  solar modulation effects for protons and nuclei. 
What concerns the electrons and positrons, the effect of solar modulation can be substantial up to 70~GeV \cite{strong11}. Therefore, for positrons  we do not expect perfect fits at energies well below 100~GeV. %

To account for the flat  $e^+/p$ and $\bar{p}/p$ ratios, an  additional component of secondary antiparticles should be invoked: 
\begin{equation}
R(E)=R1(E)+R2(E) .
\end{equation}
The ratio R1(E) is for secondaries produced in the ISM (``ISM-component"); see  Eq.(1). The ratio  R2(E) corresponds  to the second component represented by secondary particles produced inside  the CR accelerators (``S-component"): 
\begin{equation}
R2(E)=\frac{\frac{X_{\rm s}(E)}{m_p}S'(E)/N_p(E)}{(1+\sigma_{s}\frac{X_{\rm ISM}(E)}{m_p})(1+\sigma_{s}\frac{X_{\rm s}(E)}{m_p})},
\label{eq:r2}
\end{equation}
where
\begin{equation}
S'(E)={\int_E\frac{d\sigma_{p-s}(E,E')}{dE'}Q_{s}(E')dE'\tau(E)} .
\end{equation}
Here $Q_{s}(E)$ is the primary injection spectrum from the source with a power law index $\gamma_{\rm s}$,  $X_{\rm s}$ is the “grammage” accumulated inside the source. We represent $X_{\rm s}$ in the form of Eq.(5)  but with a different power-law index, $\delta_{\rm s}$.   The “ISM-component” contains the secondaries produced in the ISM by primary CRs from both the 1st and 2nd CR populations, while “S-component” is contributed only by the 2nd source population. We assume that the grammage accumulated by the  1st CR population inside the sources is not substantial, thus in the calculations below this component is neglected. 

The key point of the proposed model is the existence of the  2nd population of antiparticles produced inside the CR accelerators and then injected into ISM with hard energy spectra. 
The formation of energy distributions of the secondary particles inside the source depends on the specifics of the processes of acceleration, escape, and interactions of accelerated particles.  
Assuming that the leakage of particles from the accelerator is characterized by the 
energy-dependent escape time,
$t_{\rm esc} \propto E^{-\delta_{\rm s}}$, the spectrum of primaries established inside the source keeps to be a power-law but with the index $\gamma_{\rm s}+\delta_{\rm s}$. 
The production spectra of secondary nuclei and antiparticles have the same power-law index. 
Generally, at different epochs of the source evolution, the spectra of secondaries injected into the ISM may differ from their production spectra.  However, the injection spectrum of secondaries integrated over the accelerator’s lifetime should be close to the production spectrum with the index  $\gamma_{\rm s}+\delta_{\rm s}$.

\begin{figure*}
\includegraphics[width=0.45\linewidth]{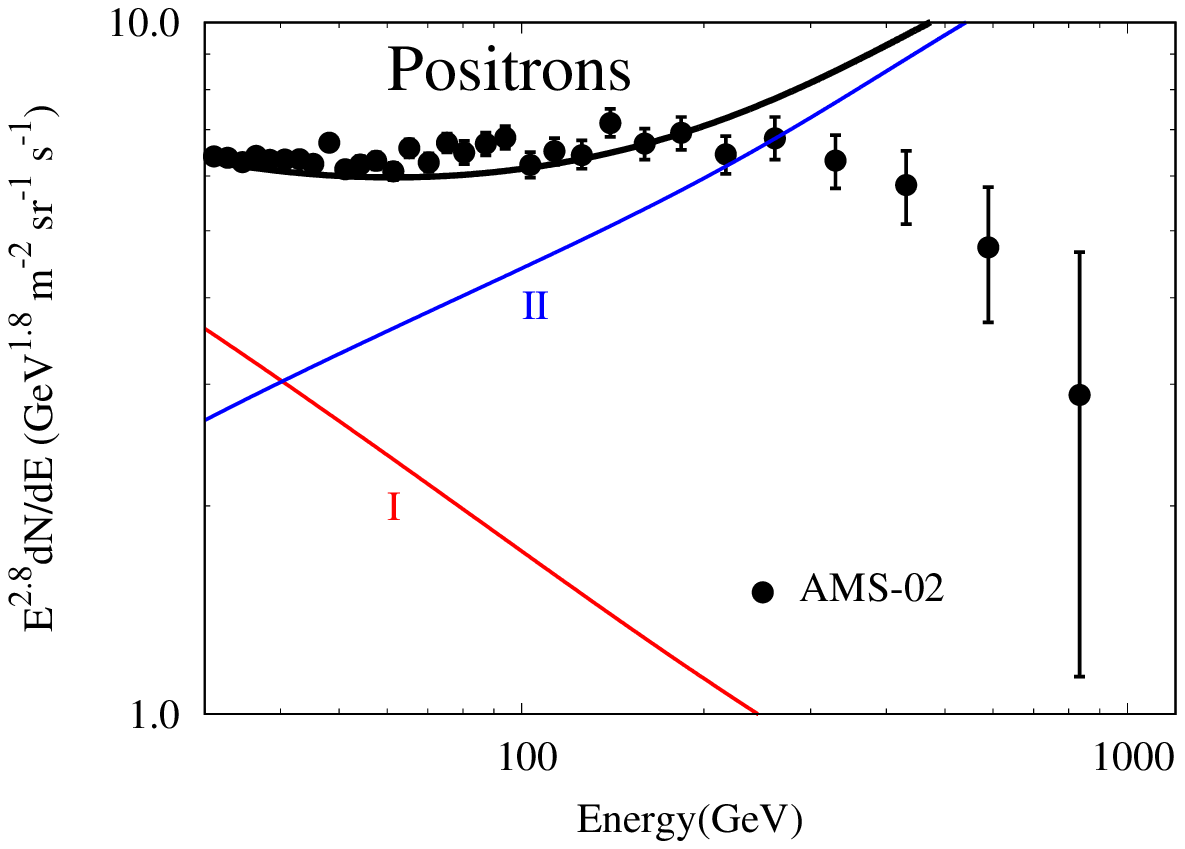}
\includegraphics[width=0.45\linewidth]{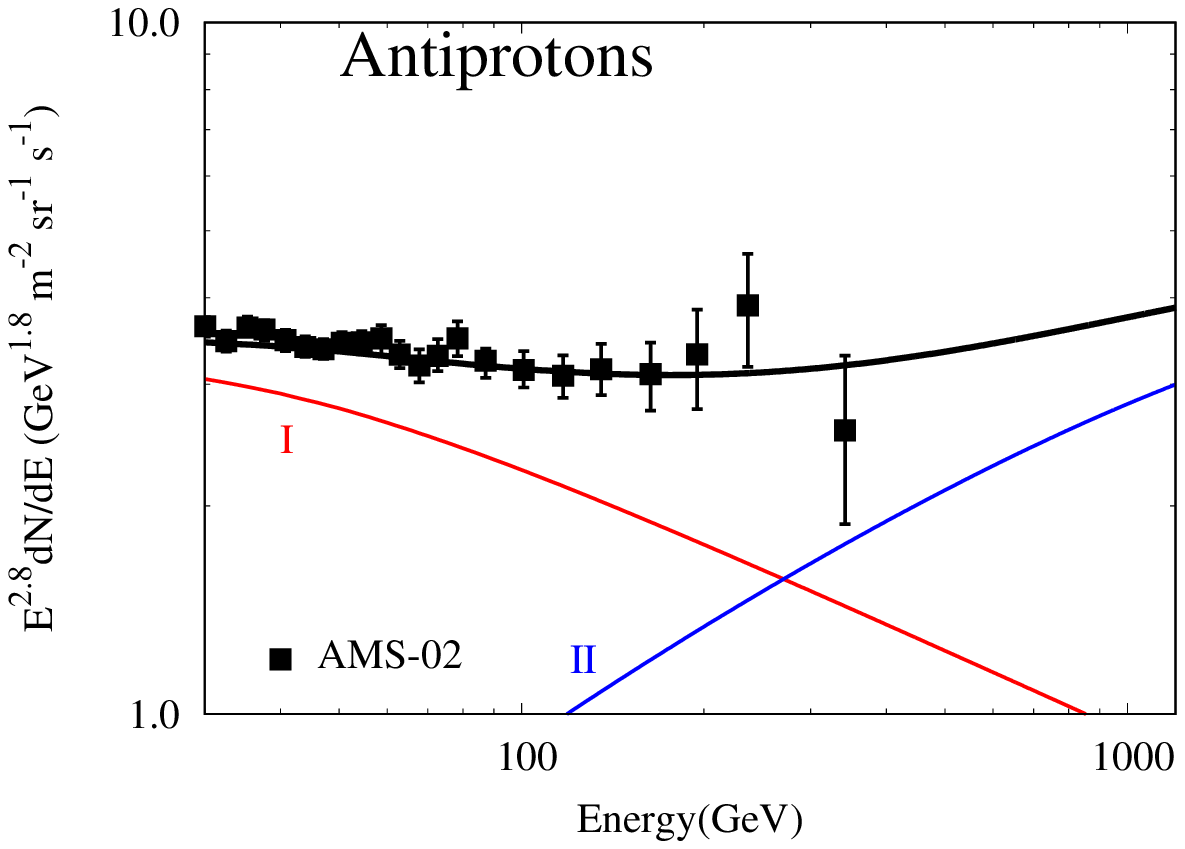}
\caption{The contributions of the two components to the detected fluxes of CR positrons and antiprotons produced in the interstellar medium and inside the sources, respectively. The black lines represent the total fluxes.  It is assumed that the power-law spectrum of parent protons from the 1st source population 
continues without a break or a cutoff,
while the spectrum of protons from the 2nd  population contains a cutoff at 100~TeV.
The data for the positron and antiproton fluxes and the antiproton/proton  ratio  are from ref.\cite{ams02posi19, ams02antip}. The points for the positron/proton ratio are derived from the data of ref.\cite{ams02posi19, ams02proton}.}
\label{fig:sec}
\end{figure*}

\begin{figure*}
\includegraphics[width=0.45\linewidth]{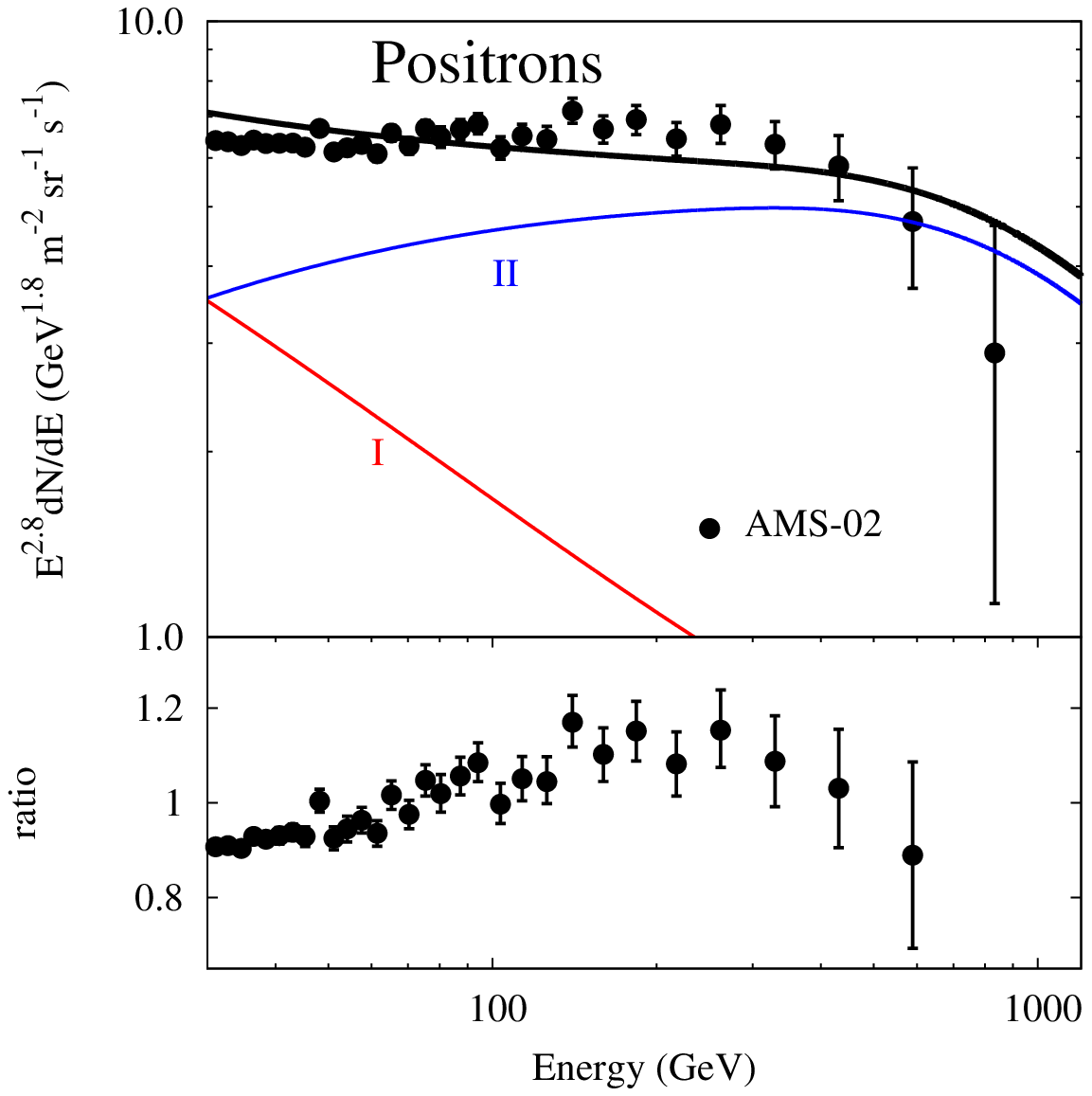}
\includegraphics[width=0.45\linewidth]{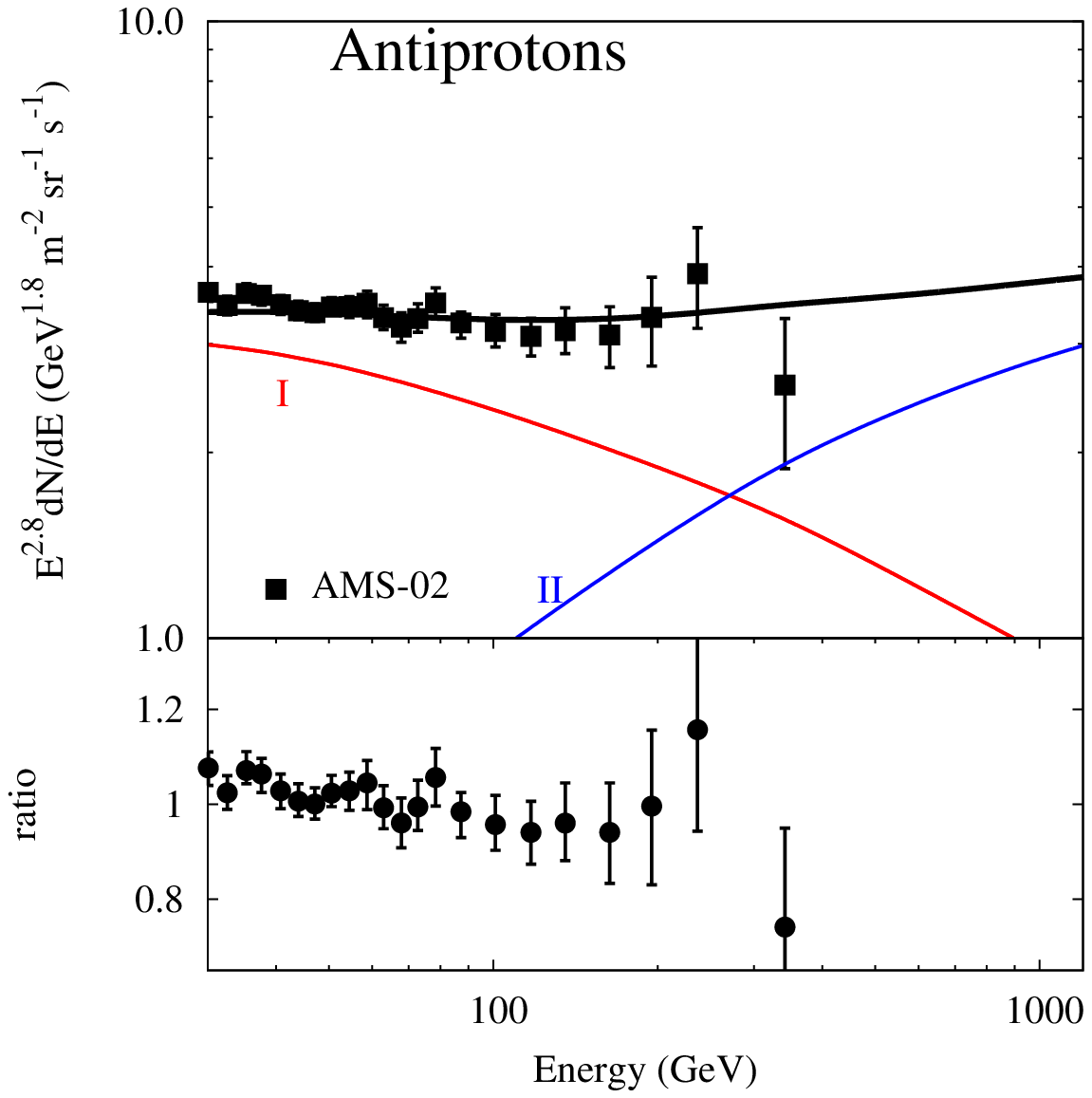}
\caption{The same as Figure \ref{fig:sec}, but assuming that an exponential cutoff in the parent 
proton spectrum at 10 TeV. Also,  to fit the data, here we assume slightly 
more grammage accumulated inside the source,  namely, 
$X_{s}=1.0~(E/\rm 10 \, GeV)^{-\delta_{\rm s}}~\rm g/cm^2$  instead of  
$X_{s}=0.8 (E/\rm 10 \, GeV)^{-\delta_{\rm s}}~\rm g/cm^2$ used in Fig.\ref{fig:sec}.  The lower panels show the ratio of experimentally measured fluxes of positrons and antiprotons to the theoretical curves.}
\label{fig:sec_cut}
\end{figure*}

The time-integrated injection spectrum of primary CRs into the ISM  repeats the acceleration spectra.  At low energies, the contribution of this hard-spectrum component  should be sub-dominant. Otherwise, it would contradict the observations.  We  normalize the contribution of the 2nd population of CRs to the total CR flux  as 
\begin{equation}
\kappa(E)=Q_{\rm s}(E) \tau_{\rm ISM}(E)/N_p(E)=\kappa_0~(\frac{E}{100~\rm GeV})^{2.85-\gamma_{\rm s}-\delta}  .
\label{eq:normalization}
\end{equation}

The analysis of the parameter space based on the adjustment of calculations to the reported fluxes of positrons and antiprotons indicates that  $\kappa_0$ cannot significantly deviate from the value of 0.1.  Therefore, we fix the latter at the level of 0.1, and derive the  best-fit values for the remaining principal parameters: $\delta=0.52, X_0=7.0 ~\rm g/cm^2, X_{s}=0.8 (E/\rm 10 \, GeV)^{-\delta_{\rm s}}~\rm g/cm^2$ and  $\gamma_{\rm s}+\delta_{\rm s} \simeq 1.8$. For these parameters, the overall energy contained in particles of the 2nd CR population above 10~GeV is approximately seven per cent of the total CR energy. Note that the above values of $X_0$ and $\delta$ are close, but not identical,  to the parameters derived from the conventional treatment based on the analysis of the B/C ratio 
under the assumption that the production of the secondary CRs takes place solely in the ISM (see. e.g., ref.\cite{yuan17}).

\begin{figure}
\includegraphics[width=1\linewidth]{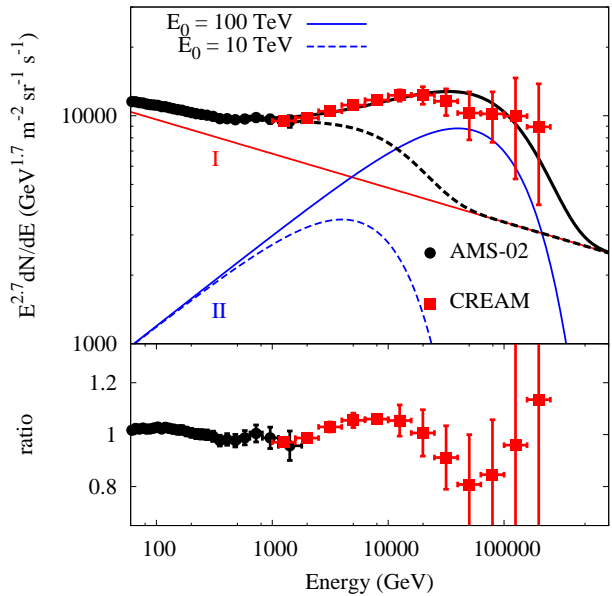}
\caption{The contributions of the 1st (I)  and 2nd (II)  source population to the proton fluxes. The black lines represent the summation of these two components. The proton data  are from the AMS-02 \citep{ ams02proton} and CREAM \cite{cream} collaborations. The spectrum of protons from the 1st source population (I) is assumed to be power law without a cutoff, while the protons from 2nd 
population (II) are calculated for two values of the exponential cutoff energy at 10~TeV (dashed) and  100~TeV (solid).  The lower panel shows the ratio of experimentally measured fluxes of protons to the theoretical curve representing the case of the cutoff in the spectrum of protons belong to the 2nd population sources at 100~TeV.}   
\label{fig:fluxpri}
\end{figure}

The results of calculations of the positron and antiprotons fluxes are shown, together with the spectral points reported by AMS-02,  in Fig.\ref{fig:sec} and  Fig.\ref{fig:sec_cut}. The agreement achieved between calculations and data robustly fixes the fluxes of both populations of primary CR protons in the ISM. Apparently, the derived proton fluxes should go along with the measurements.  The absolute fluxes of CR protons derived directly (without any normalization) from the fitting of the positron and antiproton data, indeed match quite well the direct measurements of the CR proton fluxes. It is demonstrated in Fig.\ref{fig:fluxpri}.  The flux of protons representing the 1st CR population in the ISM is described by a steep power-law spectrum, while for protons representing the 2nd population,  an exponential cutoff is introduced in the spectrum 
AT 100 TeV and 10 TeV. The corresponding fluxes of positrons and antiprotons are shown in Fig.\ref{fig:sec} and  Fig.\ref{fig:sec_cut}. 
As can be seen from these figures, for both cutoff energies, the calculations describe quite well the positron and antiproton data up to $\approx 200$~GeV. At higher energies,  the spectra of positrons and antiprotons depend on the extrapolation of the proton spectrum beyond 10 TeV. 

Note that the impact of the lower proton cutoff energy on the positron spectrum is stronger than on the antiproton spectrum.  This is explained by the fractions of the energy of the parent proton transferred to secondaries at inelastic $pp$ interactions. While the antiprotons receive more than 10 per cent of the kinetic energy of protons, the positrons receive an order of magnitude less energy. While the proton spectrum with the exponential cutoff at 100 TeV (or at higher energies) predicts a progressive hardening of the spectra of positrons and antiprotons towards 1~TeV,  the cutoff in the proton spectrum at 10 TeV results in the softening of the spectra of positrons and antiprotons. 
The latest  AMS-02 data provide strong evidence of spectral steepening of the positron flux above 200 GeV which can be interpreted as a result of the cutoff in the protons spectrum around 10~TeV (compare Fig.\ref{fig:sec} and  Fig.\ref{fig:sec_cut}).  The cutoff in the proton spectrum around 10~TeV should result in a softening of the antiproton spectrum as well; see right panel of  Fig.\ref{fig:sec_cut}). Although the last (highest energy) reported point of antiprotons shows a tendency of steepening, the statistical significance is not sufficiently high for a definite conclusion.

\begin{figure}
\includegraphics[width=1\linewidth]{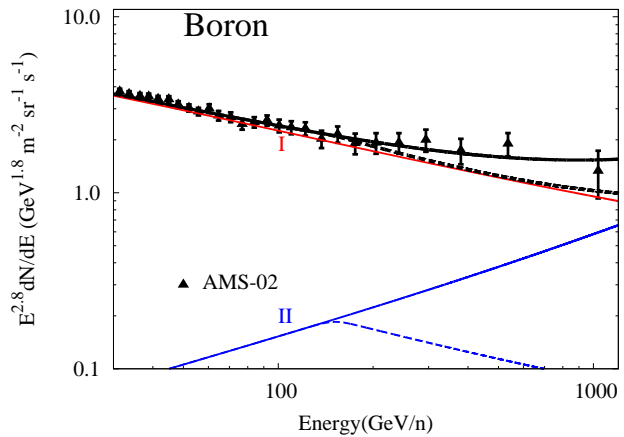}
\caption{Same as Fig.\ref{fig:sec} but for the secondary boron nuclei.  The boron fluxes are from ref.\cite{ams02sec} }
\label{fig:boron}
\end{figure}

 At interactions of primary CRs with the ambient gas, the production of secondary positrons and antiprotons is accompanied by production of secondary nuclei.  The light nuclei of the LiBeB group are of particular interest.  Because of the negligible content of these particles in the directly accelerated CRs, the measured fluxes of this group undoubtedly have a secondary origin. They are predominantly produced in interactions of primary (accelerated) nuclei belong to the CNO group. In Fig.\ref{fig:boron}, we show the fluxes of secondary boron. We should note that the good agreement between the reported fluxes and calculations in Fig.\ref{fig:boron}  is achieved assuming slightly harder injection spectra for CR nuclei compared to protons. This can be considered as a {\it ad hoc} assumption made for the interpretation of the reported fluxes of secondary light elements.  On the other hand,  this assumption perfectly agrees with the spectral measurements of primary nuclei of the  CNO group (see Fig.\ref{fig:HeOC} and discussion below).

\section{Discussion}

\subsection{Two CR components model}

In recent years we have witnessed several remarkable achievements in CR measurements.  One of the striking results concerns the high content of positrons and antiprotons in CRs. In this paper, we argue that both species are of secondary origin produced in interactions of primary  CRs with ambient matter. The absolute fluxes and the spectra of positrons and antiprotons can be naturally explained within the general framework of the current paradigm of Galactic CRs, but with two significant additions of conceptual character. First, we assume that the galactic CRs consist of  {\it two}  major components which are injected into the ISM with substantially different energy spectra. We propose that in addition to the conventional CR component with relatively steep, $FI \propto E^{-2.35}$ type initial spectrum injected into ISM, should exist the second CR component with very hard, $FII \propto E^{-1.8}$ type or harder, injection spectrum.  However, the postulation of the second CR component is not yet sufficient for interpretation of the data,  in particular, the unexpectedly  hard (for the secondary products) $E^{-2.8}$ type differential spectra of positrons and antiprotons  measured over the energy range from tens of GeV to a few 100 GeV. The second principal assumption of our model is that CRs representing the 2nd population of accelerators accumulate quite significant "grammage" of the order of 1 $\rm g/cm^2$  inside (or nearby) the sources, i.e. before their leakage into the ISM.

The spectra of parent CR protons from two source populations, derived from the fits of the antiproton and positron fluxes are shown in Fig.\ref{fig:fluxpri}.  For the 2nd CR population, 
the power-law spectral index of particles injected into
the ISM, $\gamma_{\rm s} + \delta_{\rm s}\simeq 1.8$, is required. It  can be realized by different combinations of $\gamma_{\rm s}$ and  $\delta_{\rm s}$. For example, in the case of energy-independent escape, i.e. $\delta_{\rm s}=0$, the index of the acceleration spectrum is $\gamma_{\rm s}=1.8$ and the best-fit energy-independent grammage inside the source is  $X_{s0}=0.8 ~\rm g/cm^2$. For the Kolmogorov type diffusion of particles inside the source, $\delta_{\rm s} \sim 1/3$, thus $\gamma \sim 1.5$. 

Such a spectrum is significantly harder than the acceleration spectrum predicted by the standard diffusive shock acceleration theory applied to young SNRs. Nevertheless,  SNRs, as suppliers of the 2nd CR component, cannot be excluded. In particular, it could be caused by the additional acceleration of secondaries by the same shock responsible for the acceleration of primaries.  In this context, different scenarios have been discussed. The secondaries could be produced inside the shell,  i.e. in the same region where the primary CRs are accelerated \citep{blasi_anti1,Mertsch14}, or at the penetration of accelerated protons into the dense gas clumps upstream the shock \cite{Mischa16}. In both scenarios, positrons are injected into the ISM with significantly harder spectra than the primary CRs. Alternatively, the 2nd  CR source population might be represented by young stellar clusters \cite{ysc} where particles can be effectively accelerated with exceptionally hard spectra extending to PeV energies \cite{Bykov19}.

\begin{figure}
\includegraphics[width=1\linewidth]{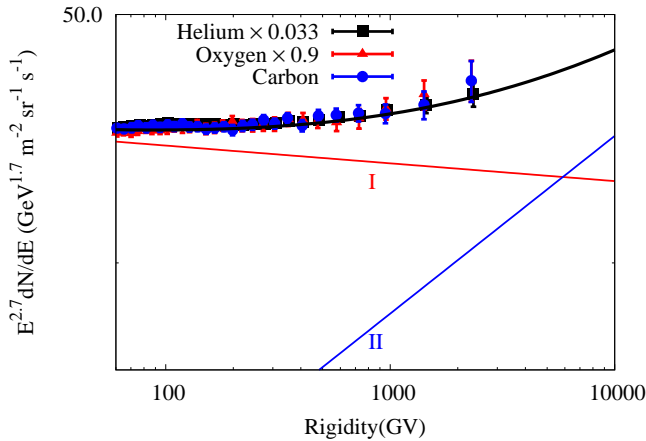}
\caption{The contributions of the 1st and 2nd source populations to the fluxes of He, C and O nuclei. The He, C, and O data are from AMS-02 \cite{ams02pri}. The fluxes of He and O are re-scaled to match the fluxes of C by the factors of 0.033 and 0.9, respectively. }
\label{fig:HeOC}
\end{figure}

While in the "secondary acceleration" models, the primaries and secondaries are injected into ISM with substantially different spectra, in our model the spectra of secondaries is only slightly harder than the energy distribution of primary CRs caused by the behaviour of production sections.  For this reason,  the postulation of the very hard primary CR component supplied by the 2nd population of particle accelerators offers a natural explanation of another firmly established feature - the effect of spectral hardening of the flux of CR protons above 200 GeV \cite{pamelap,cream, atic, ams02proton}.
Fig.\ref{fig:fluxpri} demonstrates that the superposition of these two primary CR components can reproduce the spectral hardening of protons at the transition between 100 GeV and 1 TeV. Similar behaviour with a spectral hardening around 200 GV has also been found for the CR nuclei \citep{ams02pri}. 
However, the measurements revealed that the spectra of nuclei are noticeably harder than the proton spectrum. Using the same normalization parameter $\kappa_0=0.10$ but harder injection spectra for  nuclei, namely  2.25 (instead of 2.35 for protons) for the 1st population and 1.7 (instead of 1.8 for protons)  for the 2nd  population of CRs, we can reproduce  the observed fluxes of primary CR nuclei He, C and O (Fig.\ref{fig:HeOC}) and the secondary light nuclei like  Boron (Fig.\ref{fig:boron}).   The difference in the acceleration spectra of protons and nuclei most likely is caused by the details of the injection process (see, e.g., ref.\cite{hanusch19}). 

Although in the proposed two-populations model both CR components  play important role in the formation of the energy spectra of the primary and secondary CRs, the overall energy contained in the 2nd component above 10~GeV is approximately 7 percent of the total CR energy. Thus, the introduction of the 2nd CR component does change substantially the energy requirements to the "standard" (one population) CR paradigm. Finally, we note that our model  principally differs from the two-component model proposed in ref.\cite{tomassetti15}, where for the production of secondary positrons is adopted the model of Mertsch and Sarkar \cite{Mertsch14}. The latter gives a relatively steep spectrum of protons injected into the ISM. Therefore,  for explanation of the spectral hardening of the proton spectrum around 200~GeV  the second, hard-spectrum component of CRs is postulated.
This component, unlike the 2nd CR population introduced in our model does not contribute to the production of positrons and antiprotons.

\subsection{Break in the positron spectrum?}

The superposition of two components of positrons and antiprotons produced inside CR sources with hard spectra and in the ISM  with soft spectra, predicts progressive spectral hardening at energies above several 100~GeV where the hard S-component starts to dominate over the soft ISM-component. On the other hand, the recent report of the AMS-02 collaboration  \cite{ams02posi19} shows just the opposite tendency -  a break in the positron spectrum (see Fig.\ref{fig:fluxpri}).  Different reasons might cause the steepening of the positron spectrum at highest energies. 
 
As demonstrated in Fig.\ref{fig:sec_cut}, the introduction of a cutoff in the spectrum of  2nd population of CR protons at 10 TeV (see Fig.\ref{fig:fluxpri})  can satisfactorily explain the steepening in the positron spectrum. On the other hand, at energies tens of TeV, this assumption implies a noticeable "deficit" in the proton flux compared to the measured one. To explain the observed fluxes of protons from  10 TeV to hundreds  TeV,
one has to invoke a new (3rd)  component of CRs.    While this could be a viable option, one can still avoid the introduction of the 3rd CR component, assuming that the break in the spectrum of positrons is caused by the sharp reduction of the confinement time of multi-TeV protons inside the CR accelerators. This would result in the decrease of the "grammage" accumulated by accelerated particles with an impact on the efficiency of production of secondaries leading to the steepening of the spectrum of positrons above a few 100 GeV. Formally, in such a scenario, the cutoff energy in the proton spectrum can be arbitrary high, and, therefore,  the 3rd CR component becomes redundant. 
 
The above two scenarios are relevant both to positrons and antiprotons. Although the effect of spectral steepening of antiprotons is less pronounced,  it should be revealed by future measurements at higher energies, around 1~TeV and above. Otherwise,  one should explore other explanations applicable only to positrons. In particular, the radiative losses of positrons break the compatibility between positrons and antiprotons. In this context, two different scenarios, both based on the energy-dependent losses of positrons with energy 
(${\rm d}E/{\rm d}t \propto E^2$),  could be realized:     

(i) {\it Energy losses of positrons inside the accelerators.} This scenario requires quite large magnetic field. In particular, for a supernova remnant of age $10^4$~yr, a  break in the positron spectrum would appear below several 100~GeV,  if the magnetic field exceeds 30 $\mu \rm G$.

\begin{figure}
\includegraphics[width=1\linewidth]{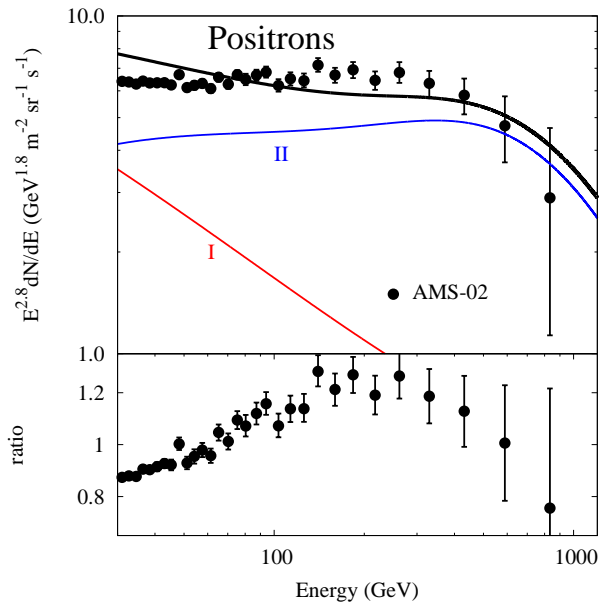}
\caption{The predicted positron spectrum with a parent proton spectrum of $N(E) \sim E^{-1.5}\rm exp(-\frac{E}{200~\rm TeV})$. The nearest source distance is assumed to be 4kpc.   The lower panel shows the ratio of experimentally measured fluxes of positrons to the theoretical curve.} 
\label{fig:posi_cool}
\end{figure}

(ii)  {\it Spatial distribution of sources}. If the sources of S-component of positrons are located at distances beyond 1~kpc, a spectral steepening caused by synchrotron and inverse Compton losses during the propagation of electrons in the ISM, is expected  (see, e.g.  ref.\citep{AAV1}). Fig.\ref{fig:posi_cool} demonstrates this possibility assuming that all sources are located at distances beyond 4~kpc.  For the achievement of a reasonable agreement with data, the spectrum of positrons injected into the ISM should be very hard to compensate for the spectral steepening in the ISM due to the radiative cooling.  Correspondingly, the spectrum of parent protons should be unusually (in particular, for SNRs)  hard with a differential power-law index $\alpha \leq 1.5$. This kind of spectra of CRs are expected in compact stellar clusters at particle acceleration caused by interactions of core-collapse supernova 
blast waves with fast winds of massive young stars \cite{Bykov19}.

\subsection{Highest energy protons}

While the maximum energy of accelerated protons is not a critical parameter for the range of energies of detected positrons and antiprotons,  the value of $E_0$ exceeding 100 TeV is needed for the explanation of the proton fluxes reported by the CREAM collaboration \cite{cream}. The distinct signature of this model is the formation of a hard spectrum of the 2nd component of protons in the ISM extending into the multi-TeV region where the second component starts to dominate over the first component. 
Therefore, in the case of lack of a break in the spectrum of the second CR  component to  1~PeV, we should expect a tendency of continuation of hardening of the proton spectrum until it achieves a single power-law form close to $E^{-2.3}$. However, Fig.\ref{fig:fluxpri}  shows that the CR proton measurements reported by the CREAM collaboration \cite{cream} do not support such spectral behaviour.  Despite the significant statistical uncertainties of these measurements above 100 TeV,  it is quite clear that one should assume a cutoff in the proton spectrum well below 1~PeV to avoid the excess of the proton flux.  

In Fig.\ref{fig:fluxpri},  we compare the predictions of the proposed two-component scenario with the CR proton measurements reported by the AMS-02 and CREAM collaborations. Note that the combination of the first and second CR  populations is robustly fixed. They are derived from the fits of the positron and antiproton fluxes shown in Fig.\ref{fig:sec}. In particular, the power-law indices of the two proton components are $\alpha_1\approx2.8$ and $\alpha_1 \approx2.35$ (the power-law indices of the injection spectra should be 1.8 and 2.35, respectively).

In left panel of Fig.\ref{fig:fluxpri} we assume pure (i.e. without high energy cutoff)  power-law spectrum for the first proton component and two spectra for the second proton component with exponential cutoffs at $E_{0}=100$~TeV and 10~TeV.  The cutoff at 100~TeV provides a good fit to the CREAM data. Despite the uncertainties in the reported proton fluxes,  especially at $E_0 \sim 100$~TeV,  the conclusion regarding the cutoff around 100~TeV is quite robust. This value, however, could be underestimated, by a factor of 2 or so, given the lack of information about the contribution of the 1st component to protons at multi-TeV energies.  In particular, for the cutoff in the 1st proton component at 10 TeV,  the corresponding suppression of the proton flux can be compensated enhancing the contribution of the second component by the increase of the cutoff energy in the 2nd component to $E_0 \approx 200$~TeV. Further increase of the cutoff energy would lead to the conflict with observations,  but for a firm conclusion, new precise measurements in the energy band $\geq 100$~TeV are needed.

\begin{figure*}
\includegraphics[width=0.45\linewidth]{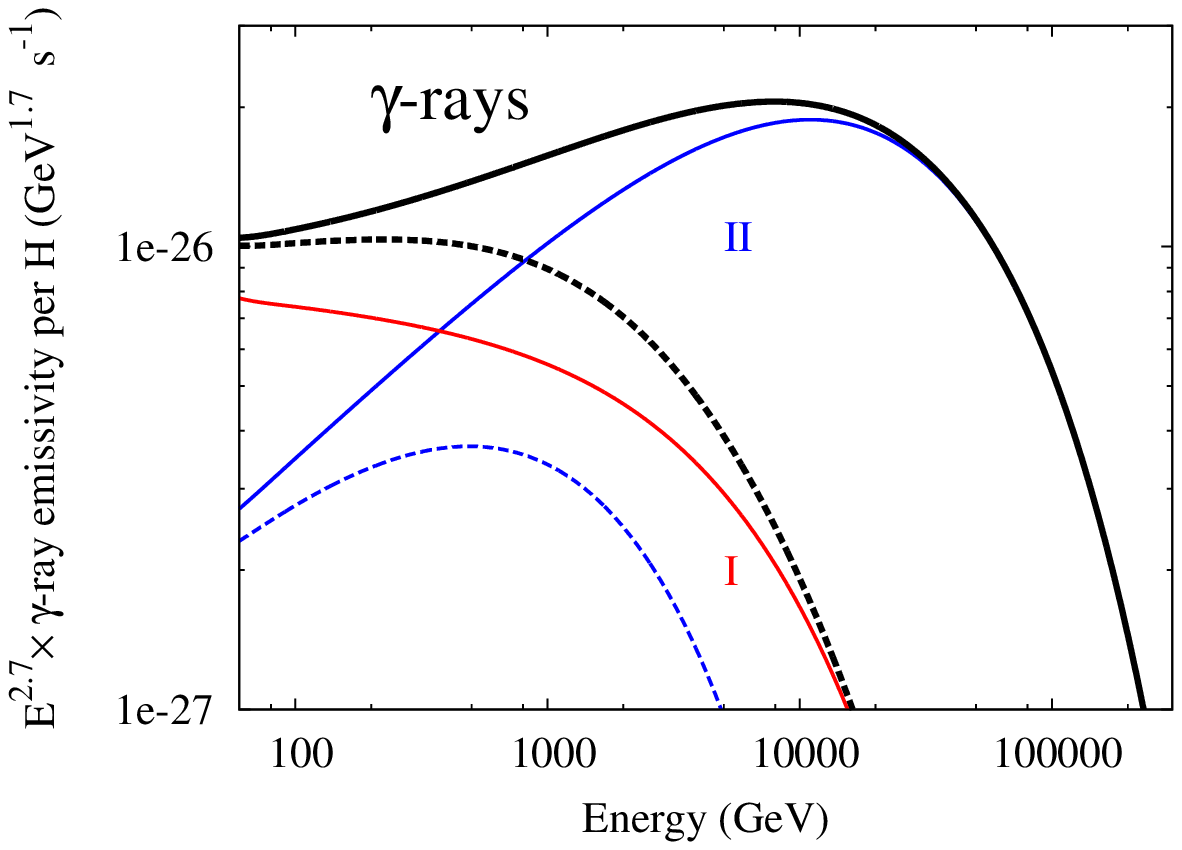}
\includegraphics[width=0.45\linewidth]{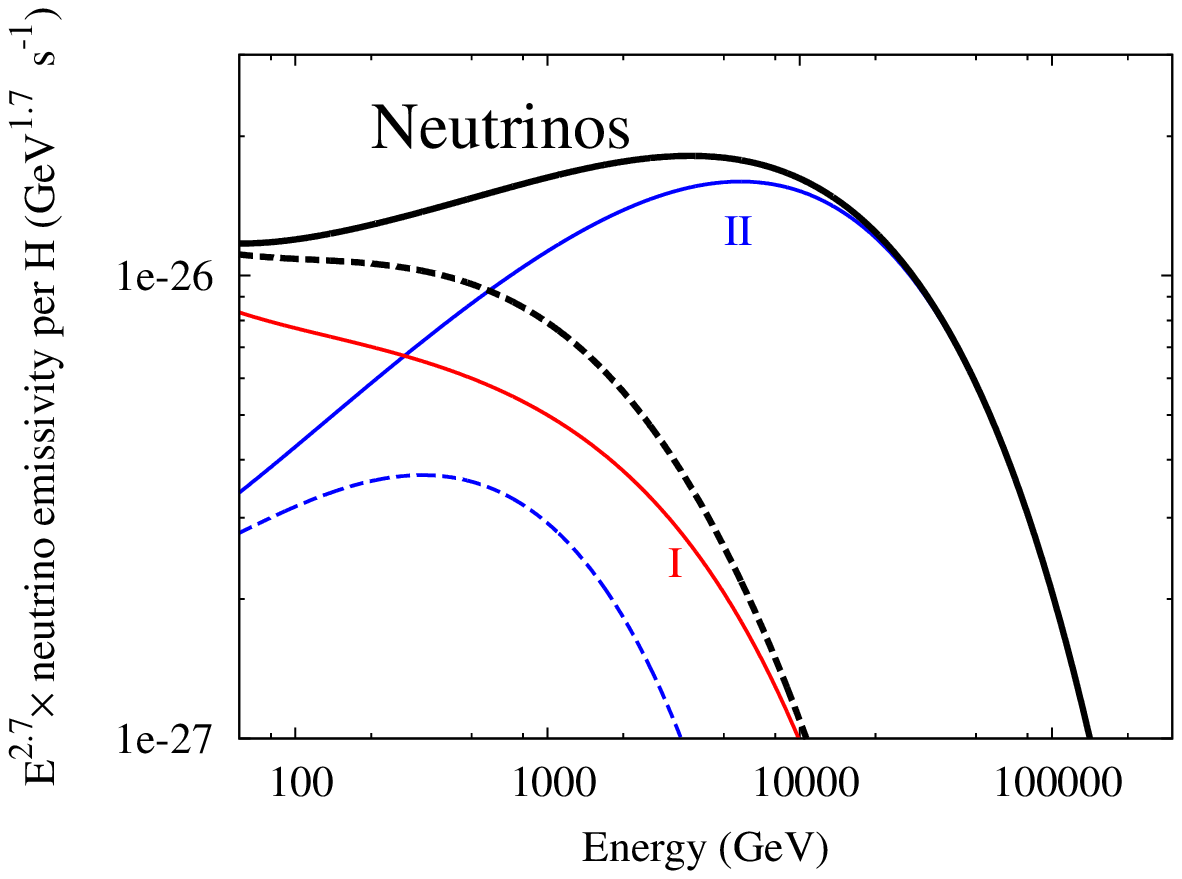}
\caption{The calculated diffuse $\gamma$-ray and neutrino emissivities contributed by CRs from 1st (red curves)  and 2nd (blue)  source populations. The protons from 1st source population are assumed to be a power-law spectrum with an exponential cutoff at 50 TeV. The spectra of protons from the 2nd source populations  are calculated for two values of the exponential cutoff energy:  1 \ PeV  (solid) and \ 0.2 \ \rm PeV (dashed). The total fluxes of gamma-rays and neutrinos are shown by black curves.}
\label{fig:fluxdif}
\end{figure*}

 Although the discussion of the nature of the postulated two CR source populations is beyond the scope of this paper, SNRs seem to be quite apparent candidates for the  1st population.  The relatively soft energy spectra of  CRs inside young SNRs with a power-law index around 2.3-2.4, as derived from $\gamma$-ray observations of SNRs, match well the observed steep energy spectrum of CRs and the energy dependence of the confinement time of CRs $\propto E^{-0.5}$ derived from measurements of the secondary nuclei.  A subgroup of  SNRs (not yet detected in $\gamma$-rays) with specific internal and external conditions could, in principle, be responsible also for the 2nd component.  Alternative sources accountable for the 2nd CR population could be clusters of young, luminous stars and associated superbubbles.   Their extended diffuse $\gamma$-ray structures recently revealed around some of these objects,   can be interpreted as a result of interactions of   CRs injected by these systems with the surrounding gas \cite{ysc}.  The injection spectrum of protons is expected to be harder than $E^{-2}$.  Furthermore, due to the long (1 Myr or more) propagation time in dense regions surrounding young stellar clusters, CRs can accumulate sufficient grammage to explain the excess fluxes of positrons and antiprotons.  Future observations of SNRs and stellar clusters with the advanced gamma-ray detectors can shed light on the origin of these two potential CR factories.
 
  We should note that in our treatment, we do not consider the possible contributions from a nearby source(s). As long as the distances to these objects are smaller than other characteristic scales, like the mean propagation paths of particles, the nearby sources principally cannot be considered as a part of the ensemble of continuously distributed sources,  but need special treatment (see, e.g. \cite{AAV1,20TeVelectrons}).  The contributions of nearby CR accelerators in certain energy bands of the CR spectrum could be nonnegligible.  In particular, the  extension of the CR electron spectrum to TeV energies provides evidence for the existence of an electron {\it Tevatron} in our neighbourhood located at distances not much further  100~pc \cite{AAV1}.  The positron fraction in the electron spectrum never achieves the level of  50 \%;  moreover, it drops above several 100~GeV. This very fact rules out the dominance of the secondary origin of multi-TeV electrons. The origin of the local electron accelerator remains highly unknown and challenging \cite{20TeVelectrons}.  The discussion of this issue is outside of the scopes of this paper.  We only note that most-likely we deal with a very efficient accelerator of primary multi-TeV electrons.  Whether this accelerator has a link to the first or second CR source populations or stands in its own right, this is an exciting issue to be explored in future work.

  \subsection{On the accuracy of spectral fits}
  
 The two-component model proposed in this paper, satisfactorily explains the general spectral features of both the primary and secondary cosmic rays: Positrons, antiprotons, secondary light nuclei like Boron,  the primary protons, as well as, with an additional assumption, the primary nuclei of the CNO group and $\alpha$-particles.  As one can see from Figs \ref{fig:sec_cut}, \ref{fig:fluxpri} and \ref{fig:posi_cool}, the accuracies of fits for positrons, antiprotons and protons are within 20 \%. Similar accuracies are achieved in the fits of the primary and secondary nuclei. The further improvement of fits of several independent measurables is not easy given the unprecedented accuracies of the measurements achieved by AMS-2 and other experiments over the recent years. We still can reduce the difference between the model predictions and the flux measurements down to 10 \% or so. However, we think that the further improvement of fits would be redundant and misleading, given the large uncertainties associated with the distribution of sources, the acceleration spectra of primary particles, the energy-dependent confinement times both inside the accelerators and in the galactic plane, etc.  

This especially concerns the positrons. As mentioned above, the impact of the solar modulation on the flux of positrons can extend up to 70~GeV \cite{strong11}.   Also, positrons suffer significant radiative losses; therefore, the calculations are sensitive to the spatial distribution of sources, the confinement of positrons inside the sources, etc. Finally one should note that although the detected fluxes of antiprotons tell us that the major fraction of positrons is of a secondary origin produced in interactions of primary cosmic rays with the surrounding gas, one cannot entirely neglect the contribution of primary positrons.  In this regard, the deficit in secondary positrons in Figs \ref{fig:sec_cut} and \ref{fig:posi_cool} can be compensated by an additional component, for example, positrons produced by pulsars/pulsar wind nebulae. Given these circumstances, an attempt of improvement of spectral fits would be, to a certain extent,  meaningless in the sense of misinterpretation of the fits as an indicator of "goodness" of the model. 

On the other hand, the large statistical errors of the measurements of the highest energy fluxes of all species of CRs are not sufficient for making certain conclusions regarding the realization of different scenarios. This concerns, in particular, the fluxes of antiprotons above 200 GeV. If the new measurements will not reveal a cutoff, the already detected break in the positron spectrum would be referred to the radiative energy losses of positrons in the interstellar medium or inside the sources. Otherwise, the reason of the break in the positron spectrum would be either the cutoff in the primary proton spectrum around 10 TeV or can be explained by the fast escape of highest energy protons from the source reducing the "grammage" accumulated by protons at highest energies. The current data provided by the CREAM measurements prefer the second option (see Fig.\ref{fig:fluxpri}). But for robust conclusions, we need higher quality data of proton fluxes above 100 TeV.

 \subsection{Diffuse gamma rays and neutrinos}
 
  The intrinsic feature of the proposed model is the enhanced gamma-ray and neutrino fluxes caused by interactions of  the second CR population with the ambient gas in the vicinity of sources.   
  Due to the expected hard $\gamma$-ray spectra extending beyond 10~TeV, the CR factories responsible for the 2nd CR populations, could be possible targets for observations with the next generation $\gamma$-ray detectors such as CTA \cite{cta} and LHAASO \cite{lhaaso}. The chances of detection of these sources depend on several factors including the power and duration of active phases of particle acceleration in these objects, the number of currently operating accelerators, {\it etc}. A more definite conclusion can be drawn for the diffuse TeV $\gamma$-ray emission linked to this component. The $\gamma$-ray emissivities initiated by the 1st and 2nd CR populations in the ISM are shown in left panel of Fig.\ref{fig:fluxdif}. Above 10 TeV the 2nd component exceeds, by order of magnitude,  the $\gamma$-ray  flux related to the 1st component. The fluxes of accompanying multi-TeV neutrinos are also significantly enhanced compared to the 1st component (see Fig.\ref{fig:fluxdif}b), dramatically increasing the chances to be detected by the km-cube scale water or ice neutrino detectors.

\bibliography{antip}
\end{document}